\documentclass[preprint]{revtex4-1} 

\usepackage{graphicx} 
\usepackage{hyperref}

\begin{document}

\title{Theoretical studies of equilibrium beta limit in heliotron plasmas}

\author{Y.~Suzuki}
\affiliation{National Institute for Fusion Science, Toki 509-5292, Japan}
\affiliation{SOKENDAI, The Graduate University for Advanced Studies, Toki 509-5292, Japan}
\author{K.~Y.~Watanabe}
\affiliation{National Institute for Fusion Science, Toki 509-5292, Japan}
\affiliation{Graduate School of Engineering, Nagoya University, Nagoya 464-8603, Japan}
\author{S.~Sakakibara}
\affiliation{National Institute for Fusion Science, Toki 509-5292, Japan}
\affiliation{SOKENDAI, The Graduate University for Advanced Studies, Toki 509-5292, Japan}
\author{N.~Nakajima}
\affiliation{National Institute for Fusion Science, Toki 509-5292, Japan}
\author{N.~Ohyabu}
\affiliation{National Institute for Fusion Science, Toki 509-5292, Japan}

\email{suzuki.yasuhiro@nifs.ac.jp}

\begin{abstract}
In this study, the MHD equilibrium limit of LHD plasmas is studied using a 3D MHD equilibrium calculation code, HINT, which is an initial value solver based on the relaxation method without the assumption of nested flux surfaces. For finite beta equilibria, flux surfaces become stochastic in the peripheral region. However, though the axis shifts until the conventional limit, the separatrix to limit the equilibrium does not appear. For high beta equilibria, the force balance starts breaking in the stochastic region. To keep the force balance, the pressure gradient in the stochastic region decreases and the fixed profile is reduced. As the result, the volume averaged beta saturates. The beta value, where the force balance starts breaking inside the stochastic region, is proposed as an index of equilibrium beta limit in heliotron plasmas.
\end{abstract}

\maketitle

\section{Introduction}

Generating and keeping good magnethydrodynamic (MHD) equilibrium are aims of magnetic confinement researches. Usually, MHD equilibrium degrades due to increasing $\beta$ and then reaches the equilibrium beta limit. In the conventional theory, the equilibrium beta limit is defined by the Shafranov shift $\Delta/a$, where $\Delta = (R_{ax}(\beta)-R_{ax}(0))$ and $a$ is the effective minor radius. If the Shafranov shift achieves about 0.5, the equilibrium is limited. For tokamaks, at the equilibrium beta limit the separatrix appears, because the poloidal field is canceled by the external vertical field to keep the MHD equilibrium~\cite{Miyamoto2005}. The separatrix in the core leads to the degradation of the confinement. In stellarator/heliotron, the equilibrium beta limit can be defined by the Shafranov shift as well as tokamaks. However, Hayashi \textit{et al}. pointed out the equilibrium beta limit is also defined by the stochasticity of magnetic field lines and it is more severe than the limit defined by the Shafranov shift~\cite{Hayashi1990}. Since the pressure-induced perturbation breaks the symmetry of the magnetic field, the stochasticity of magnetic field lines by the equilibrium response is an intrinsic property in stellarator/heliotron. The problem how to define and represent the equilibrium beta limit is still an open question. Thus, understanding the equilibrium beta limit is a critical and urgent issue from the viewpoint to aim stellarator/heliotron reactor.

\begin{figure}[htbp]
 \includegraphics{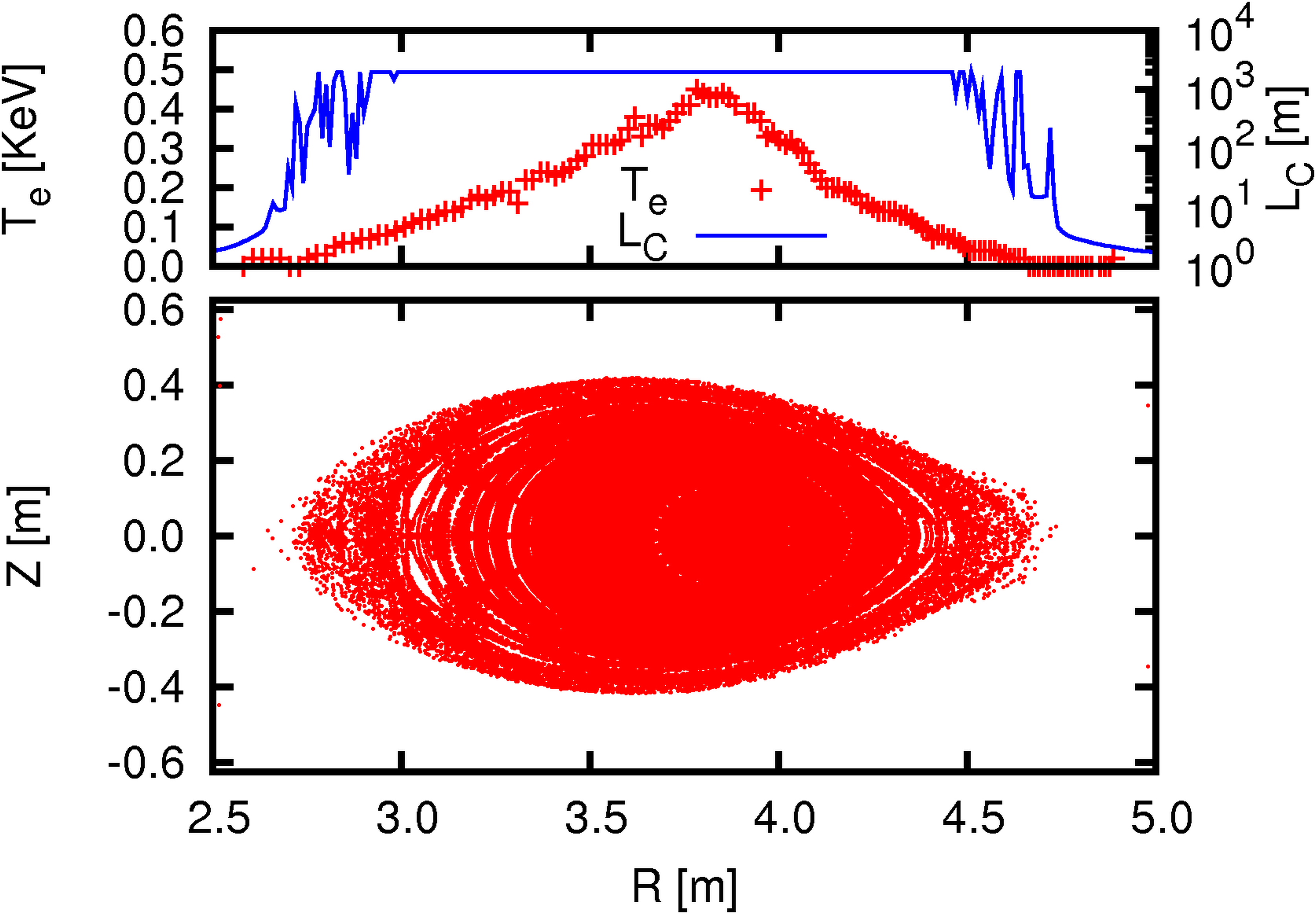}
 \caption{A Poincar\'e plot of magnetic field lines for $\langle \beta \rangle \sim$ 4.8\% is plotted at the horizontal elongated cross section. Profiles of $L_\mathrm{C}$ and $T_e$ (\#69901) are also plotted as the reference.}
 \label{fig:fig1}
\end{figure}

The Large Helical Device (LHD) is an $L$=2/$M$=10 heliotron device. In 10th experimental campaign, the beta value achieved up to 4.8\%, which is relevant to a parameter of the reactor operation, in quasi-steady state operation~\cite{Sakakibara2008}. Figure~\ref{fig:fig1} shows Poincar\'e plot of magnetic field lines for a finite-$\beta$ equilibrium best fitted to the experimental result. Profiles of the connection length $L_\mathrm{C}$ from the equilibrium calculation and the electron temperature $T_e$ from the experimental result (shot \#69910) are also shown as the reference. The magnetic axis shifts from 3.53m to about 3.8m and magnetic field lines become stochastic especially in the peripheral region. Despite of magnetic field lines becoming stochastic, it is seemed that Te spreads over in the peripheral region and the gradient of Te exists in the region. Suzuki \textit{et al.} pointed out the possibility that the stochastic field lines can keep $T_e$~\cite{Suzuki2006} in a following situation. If $L_\mathrm{C}$ is sufficiently longer than the electron mean free path (e-mfp) $\lambda _e$, the field line can keep finite temperature gradient along the field line from the viewpoint of the transport. The estimated $\lambda _e$ from the profile in the fig.~\ref{fig:fig1} is less than 10m in the peripheral region. It is much shorter than $L_\mathrm{C}$. Thus, the reconstructed equilibrium is consistent to the experimental result. However, though the beta value achieved about 5\%, the study of the equilibrium beta limit is not only important in the theory but also reactor design.

In this study, we study theoretically the equilibrium beta limit in LHD plasmas as a typical heliotron configuration. Special notice is how to define and represent the equilibrium beta limit. Analyses were done by a 3D MHD equilibrium code, HINT~\cite{Suzuki2006}, which does not assume nested flux surfaces. Using HINT, we can study not only the Shafranov shift but also the stochasticity of magnetic field lines on the equilibrium beta limit. In next section, we study finite-$\beta$ effects in heliotron plasmas. Then, we discuss how to define the equilibrium beta limit and summarize.


\section{MHD equilibrium properties in an optimized configuration for high-$\beta$ operation}\label{sc:Er}

In this section, we study MHD equilibrium properties in an optimized configuration for high-$\beta$ operation. In LHD experiments, the achieved beta value strongly depends on preset magnetic axis for the vacuum field $R_\mathrm{axV}$ and the plasma aspect ratio $A_\mathrm{p}$~\cite{Watanabe2004}. In the standard configuration, $R_\mathrm{axV}$ and $A_\mathrm{p}$ are set to 3.6m and 5.8, respectively. The achieved beta value in the standard configuration was about 3\% and the Shafranov shift approached about 0.5. Thus, in order to reduce the Shafranov shift, the optimization of $A_\mathrm{p}$ was done, because the Shafranov shift is relation to the inverse of $A_\mathrm{p}$~\cite{Motojima2007}. For $A_\mathrm{p}$=6.6, highest beta value was obtained. In this study, we study this optimized configuration.

3D MHD equilibrium calculations were done by the HINT code, which is an initial value solver based on the relaxation method. Since HINT uses a real coordinate system, which is called to the rotating helical coordinate system, it can consider unclosed flux surfaces. One advantage of HINT is the calculation including the real coil geometry, because the stochasticity of magnetic field lines in the peripheral region strongly depends on the shape of external coils. The pressure profile is prescribed to $p=p_0(1-s)(1-s^4)$. Here, $s$ is the normalized toroidal flux. In previous studies, the pressure profile was only prescribed as an initial profile. Since the HINT uses the relaxation method, initial and relaxed profiles were different in many cases. Changing the pressure profile in the relaxation, it is difficult to distinguish whether the stochasticity of magnetic field lines is driven by the plasma pressure or profile. In order to resolve that problem, we modify the code to fix the pressure profile in the relaxation. In the relaxation, $s$=1 magnetic surface always passes through $R$=4.6m, a torus outboard point on an equator defined as an input parameter.

\begin{figure}[htbp]
 \includegraphics{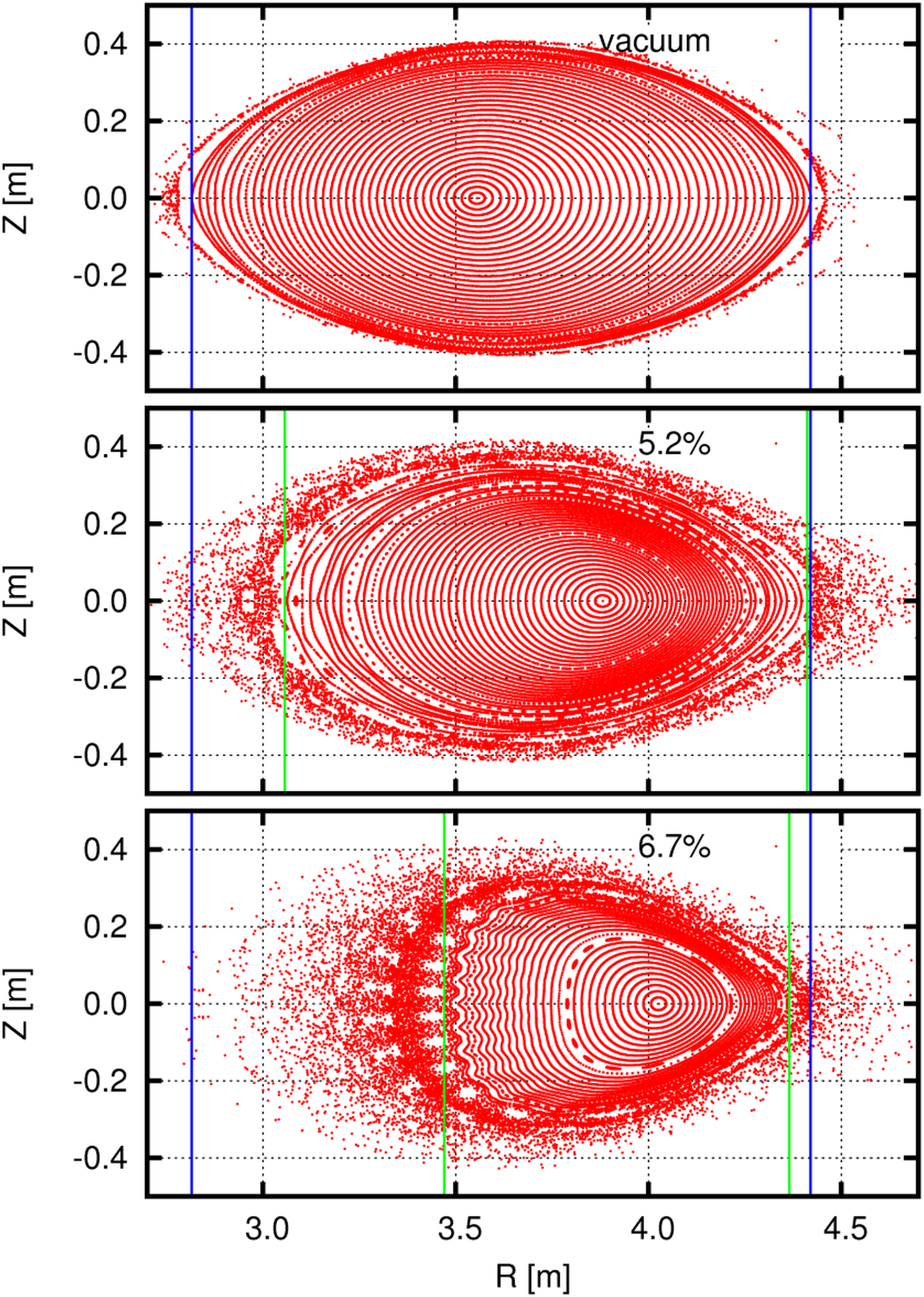}
 \caption{Poincar\'e plots of magnetic field lines are shown for finite-b equilibria. Vacuum field lines are also shown as the reference. Blue and green lines indicate position for the LCFS along Z=0 const. plane.}
 \label{fig:fig2}
\end{figure}

Figure~\ref{fig:fig2} shows Poincar\'e plots of magnetic field lines for MHD equilibria of the optimized configuration at the horizontal elongated cross section. As the reference, the vacuum field is also plotted in the same format. Blue lines in figs. indicate well-defined last closed flux surface (LCFS) for the vacuum along $Z$=0 \textit{const}. plane and green lines indicate the LCFS for the finite-$\beta$. For the vacuum field, clear flux surfaces are keeping till the peripheral region. Increasing $\beta$, the magnetic axis and the inward LCFS shift to the outside of the torus. In addition, magnetic field lines become stochastic in the peripheral region. However, for $\langle \beta \rangle \sim$5.2\%, the outward LCFS is still kept in the vacuum LCFS. For higher $\langle \beta \rangle $ ($\sim$6.7\%), the LCFS shrinks in the inside the vacuum LCFS. Increasing the stochasticity, the plasma volume with clear flux surfaces deceases about 45\% for $\langle \beta \rangle \sim$6.7\%. 

\begin{figure}[htbp]
 \includegraphics{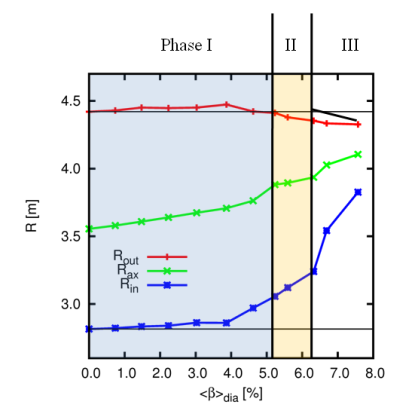}
 \caption{The change of the axis ($R_\mathrm{ax}$), inward and outward positions ($R_\mathrm{in}$ and $R_\mathrm{out}$) of the LCFS on horizontal elongated cross section (along $Z$=0 plane) are plotted as the function of $\langle \beta \rangle$. }
 \label{fig:fig3}
\end{figure}

In order to understand the stochasticity of magnetic field lines due to increasing $\beta$, positions of the magnetic axis ($R_\mathrm{ax}$), inward and outward LCFS ($R_\mathrm{in}$ and $R_\mathrm{out}$) as the function of $\langle \beta \rangle $ are plotted in fig.~\ref{fig:fig3}. The magnetic axis $R_\mathrm{ax}$ monotonically changes due to increasing $\beta$. On the other hand, though $R_\mathrm{ax}$ and $R_\mathrm{in}$ shift outward due to increasing $\beta$, $R_\mathrm{out}$ is fixed near the vacuum LCFS until $\langle \beta \rangle \sim$5.2\%. However, for higher $\beta$, $R_\mathrm{out}$ shrinks inside the vacuum LCFS and magnetic field lines become strongly stochastic. In more than $\langle \beta \rangle $=6.3\%, the force balance starts breaking inside the vacuum LCFS. To keep the force balance, the pressure gradient in the stochastic region decreases and the fixed profile is reduced with decreasing $R$ to fix the pressure profile. For $\langle \beta \rangle <$ 6.3\%, $s$=1 flux surface is a surface pass through $R$=4.6m. However, for $\langle \beta \rangle >$ 6.3\%, the stochastic field cannot keep the pressure gradient until $R$=4.6m. Thus, fixed point $R$ to prescribe the pressure profile is decreased to $R$=4.4m. As the result, total stored energy $W_p$ decreases. We summarize the MHD equilibrium in heliotron plasmas involves following three phases in fig.~\ref{fig:fig3}. (I) the plasma shifts due to the finite beta but $R_\mathrm{out}$ is fixed near the vacuum LCFS. (II) Increasing beta, the stochastic region increases and the LCFS shrinks compared with the vacuum. (III) Increasing the stochasticity, the residual force balance is broken. The magnetic configuration cannot keep the plasma.

\section{Discussions}\label{sc:HINT}

\begin{figure}[htbp]
 \includegraphics{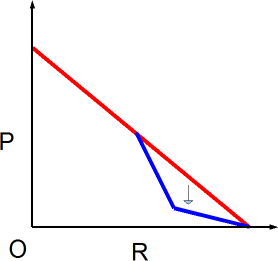}
 \caption{The schematic view of reduced pressure profile due to stochastic field lines. }
 \label{fig:fig4}
\end{figure}

\begin{figure}[htbp]
 \includegraphics{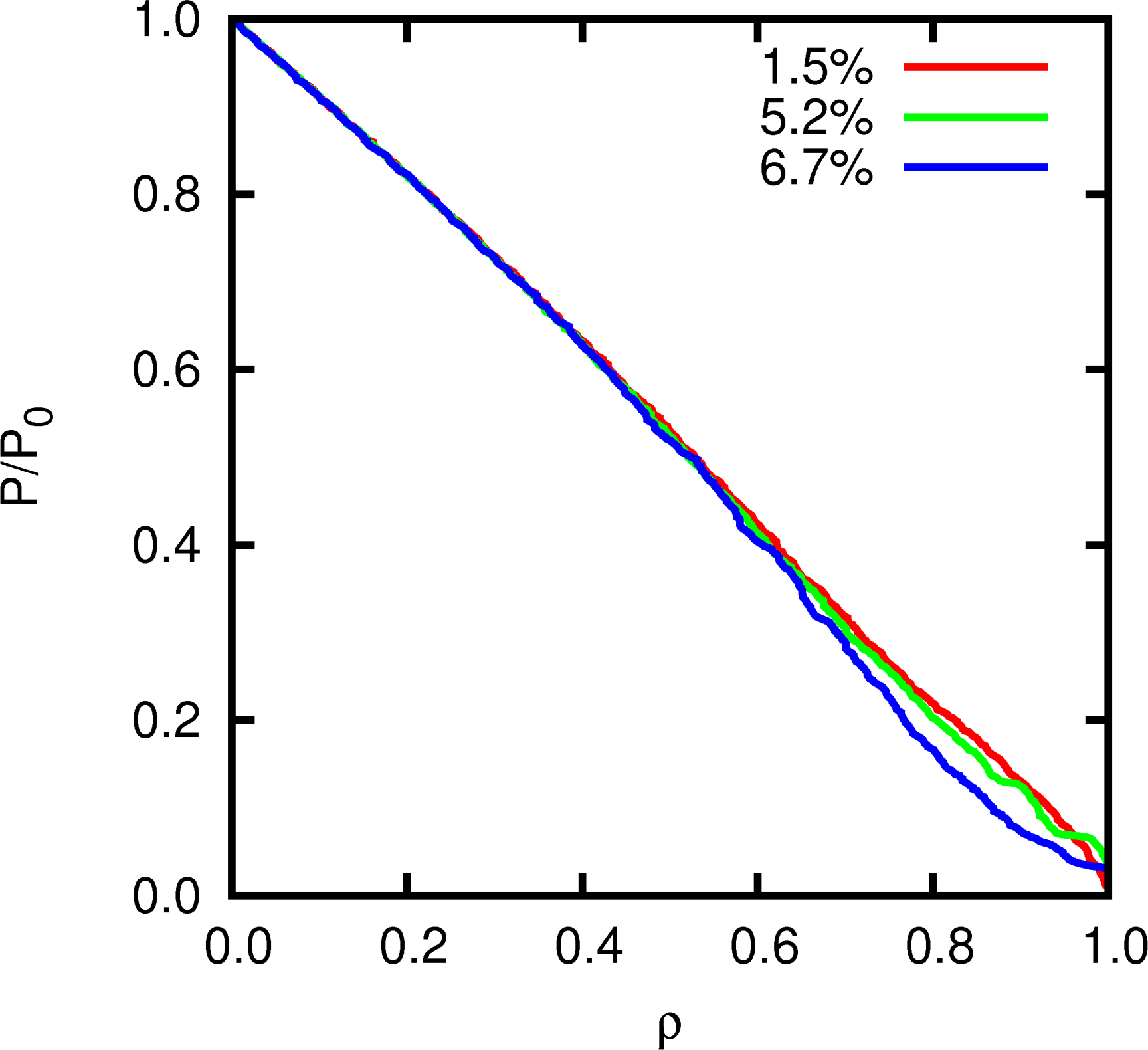}
 \caption{Profiles of the plasma pressure for various $\beta$ are plotted as the function of the normalized minor radius. }
 \label{fig:fig5}
\end{figure}

As we had mentioned, if magnetic field lines become strongly stochastic, stochastic field lines cannot keep steep pressure gradient and the fixed pressure profile is reduced. Figure~\ref{fig:fig4} shows a schematic view of the reduced pressure profile. A red line indicates a fixed pressure profile as an input parameter. If magnetic field lines do not become strongly stochastic, it can keep the pressure gradient along the red line. However, if the stochasticity of magnetic field lines is strong, the pressure gradient decreases to satisfy the force balance, $\nabla p = \textbf{j} \times \textbf{B}$. Since the pressure gradient is reduced, the total stored energy is also reduced. This represents the reduction of the volume averaged beta against the beta on the axis. In fig.~\ref{fig:fig5}, the schematic view to represent the reduction of the volume averaged beta is shown. Usually, the volume averaged beta is relation to the beta on the axis, because the pressure profile is prescribed as the function of the flux, the plasma volume and so. However, for the higher $\beta$ equilibrium, the magnetic field cannot keep the prescribed profile and the linearity inflects. In fig.~\ref{fig:fig5}, profiles of the plasma pressure obtained from the equilibrium calculation are shown as the function of the normalized minor radius. The normalized minor radius $\rho$ is defined as $\rho=\sqrt{s}$. In fig.~\ref{fig:fig5}, for $\langle \beta \rangle \le $ 6.3\%, pressure profiles are almost same. However, for $\langle \beta \rangle $=6.7\%, the profile is same to $\rho \sim$0.6 but in more than $\rho$=0.6 the profile is different. This leads the reduction of the volume averaged beta. We define this inflection points as an index of the equilibrium beta limit (between phase II and III in fig.~\ref{fig:fig3}). Thus, the equilibrium beta limit is observed as the degrading of the stored energy. Special notice is this index does not mean the saturation of increasing the volume averaged beta. In fig, only the ratio between $\beta_0$ and $\langle \beta \rangle $ is reduced and the beta does not saturate.

\begin{figure}[htbp]
 \includegraphics{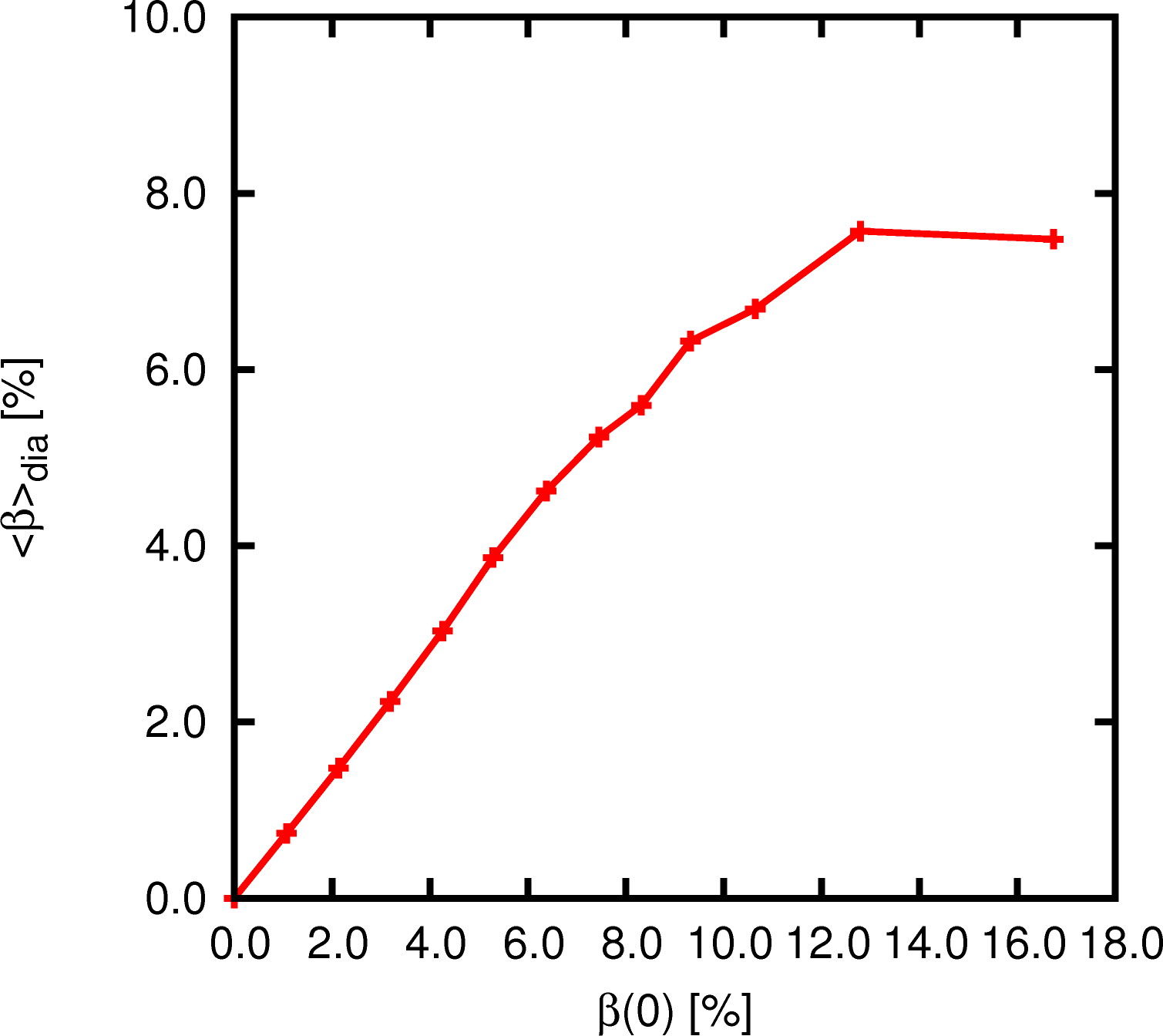}
 \caption{The volume averaged beta is plotted as the function of the beta on the axis. The linearity inflects at $\langle \beta \rangle \ge$ 6.3\% ($\beta_0 \ge$ 10\%). }
 \label{fig:fig6}
\end{figure}

In order to confirm our speculation, the volume averaged beta obtained from the calculation is plotted as the function of the beta on the axis. Figure~\ref{fig:fig6} shows the relation between $\beta_0$ and $\langle \beta \rangle $. Two auxiliary lines are plotted as the reference. For $\langle \beta \rangle \le$6.3\%, the volume averaged beta increases almost linearly. Since the pressure profile is prescribed as the function of the toroidal flux, which means the function of the cross section, it is not completely linear. However, for higher $\beta$ ($\langle \beta \rangle \ge$ 6.3\%), the stochasticity becomes strong and the fixed pressure profile is reduced to keep the force balance. Thus, an inflection points appeared for $\langle \beta \rangle \sim$6.3\%. In the considered condition, the equilibrium beta limit is more than 6\%. This mean the experimental result (\#69910) for $\langle \beta \rangle \sim$4.8\% does not achieve the equilibrium beta limit.

\begin{figure}[htbp]
 \begin{center}
  \begin{tabular}{c}
   (a) \\
   \includegraphics{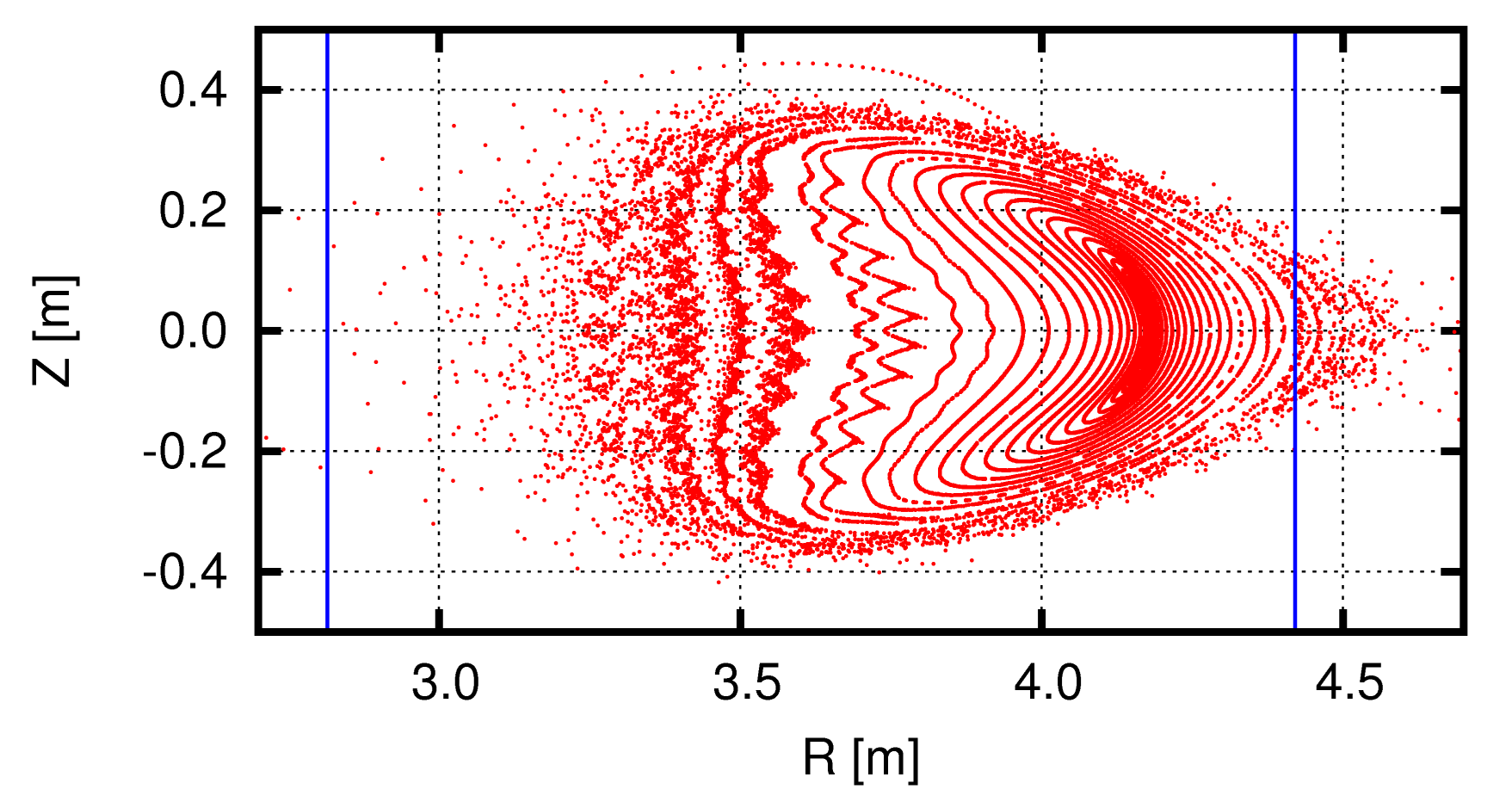} \\
   (b) \\
   \includegraphics{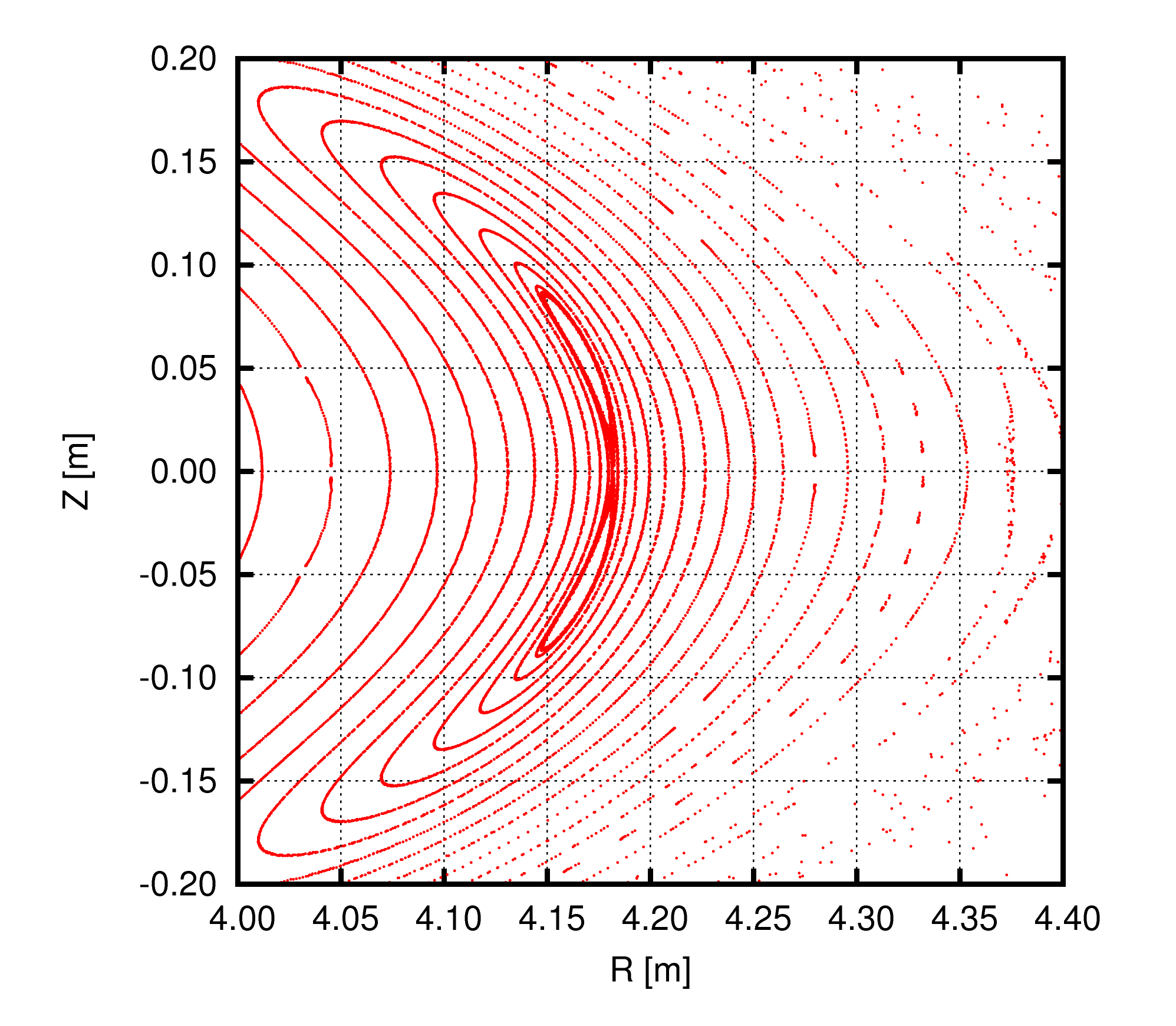}
  \end{tabular}
 \end{center}
 \caption{(a) Poincar\'e plot and (b) detailed plot near the axis of magnetic field lines and (b) are plotted for $\beta_0 \sim$ 16\%.}
 \label{fig:fig7}
\end{figure}

Increasing the stochasticity of magnetic field lines, the increasing of the volume averaged beta degrades. However, though the gradient of the function of $\beta_0$ becomes weak, the volume averaged beta keeps increasing. The reduction of the fixed pressure profile means the increasing of the peaking factor of the profile. Since the pressure profile becomes effectively peaked, the magnetic axis shifts further. In order to study the equilibrium beyond the index of the equilibrium beta limit, we study the MHD equilibrium for higher $\beta$ ($\langle \beta \rangle \ge$ 6.5\%). Figure~\ref{fig:fig7} (a) and (b) shows Poincar\'e plot of magnetic field lines for $\beta_0 \sim$16\% and detailed plot near the axis, respectively. In fig.~\ref{fig:fig7} (a), magnetic field lines become strongly stochastic. In addition, the plasma shape strongly elongates because of large Shafranov shift. Since the vertical field driven by the Pfirsh-Schl\"uter current cancels the poloidal field generated by the external coil system, the separatrix appears near the axis in fig.~\ref{fig:fig7} (b). In order to satisfy the force balance on the separatrix, the plasma pressure flatters on the separatrix. Thus, the confinement degrades because of flattering of the plasma pressure. Green circle in fig.~\ref{fig:fig6} indicates the generation of the separatrix. Though the beta on the axis increases from 12\% to 16\%, the volume averaged beta is almost constant. The generation of the separatrix gives most severe limit.

The generation of the stochastic region in the peripheral region is an intrinsic property in stellarator/heliotron plasmas. Since the transport strongly links the MHD equilibrium, it seems that the stochasticity affects the transport procces. Rechester \textit{et al}. suggested the stochasticity of magnetic field affects the electron heat transport~\cite{Rechester1978}. Thus, there is a possibility that the stochasticity limits the pressure gradient for $\langle \beta \rangle \le$6.3\% because of increasing the transport. Since such an effect is important in the reactor, it is necessary to study the reduction of the stochasticity. In addition, the net-toroidal current also strongly affects the MHD equilibrium and the stochasticity in stellarator/heliotron. In this study, we discuss only the net-toroidal current free equilibrium. Since we observe the net-toroidal current in the experiment, it is necessary to discuss the MHD equilibrium including the net-currents. Those are future subjects.


\section{Summary}

We discuss the equilibrium beta limit for a LHD configuration as a typical heliotron configuration. Increasing $\beta$, magnetic field lines become stochastic by the equilibrium response. If stochastic field lines invade in the plasma core region, the increasing of the volume averaged beta degrades because stochastic field lines cannot keep steep pressure gradient. We propose this result as an index of the equilibrium beta limit in the heliotron.


\section*{Acknowledgment}

This work is performed with the support and under the auspices of the NIFS Collaborative Research Program NIFS07KLPH004 and NIFS10KLPH006. This work was supported by Grant-in-aid for Scientific Research (C) 25420890 from the Japan Society for the Promotion of Science (JSPS).


\bibliography{pfr}

\end{document}